\documentclass[twocolumn,aps,prapplied,10pt,superscriptaddress]{revtex4-2}
\usepackage[T1]{fontenc}
\usepackage{amsmath,graphicx,amssymb,epsfig,dsfont,mathrsfs,color,amsbsy}

\begin{document}

\title{Near-field focusing and amplification of tip-substrate radiative heat transfer}

\author{M. Vescovo}
\affiliation{Laboratoire Charles Fabry, UMR 8501, Institut d'Optique, CNRS, Universit\'{e} Paris-Saclay, 2 Avenue Augustin Fresnel, 91127 Palaiseau Cedex, France}
\author{P. Ben-Abdallah}
\email{pba@institutoptique.fr} 
\affiliation{Laboratoire Charles Fabry, UMR 8501, Institut d'Optique, CNRS, Universit\'{e} Paris-Saclay, 2 Avenue Augustin Fresnel, 91127 Palaiseau Cedex, France}
\author{R. Messina}
\email{riccardo.messina@institutoptique.fr} 
\affiliation{Laboratoire Charles Fabry, UMR 8501, Institut d'Optique, CNRS, Universit\'{e} Paris-Saclay, 2 Avenue Augustin Fresnel, 91127 Palaiseau Cedex, France}

\date{\today}

\begin{abstract}
The spatially resolved near-field radiative heat transfer between a nanoscale probe and a substrate is studied in the fluctuational electrodynamics framework within the dipolar approximation. It is shown that the introduction of a thin polar film atop a non-dispersive substrate can lead to both an enhancement and a lateral focusing of the heat exchange. The influence of the probe–substrate separation, film thickness and substrate permittivity is analyzed, revealing that the effect originates from near-field interactions governed by the interplay between film-induced modifications of electromagnetic mode dispersion and the distance-dependent coupling strength. The results highlight a viable route toward the active control of local radiative heat transfer at the nanoscale.
\end{abstract}

\maketitle

\section{Introduction}

When two bodies at different temperatures are separated by a vacuum gap, heat can be exchanged between them via the emission and absorption of thermal photons. In the far-field regime--where the separation exceeds the thermal wavelength $\lambda_\mathrm{th}=\hbar c/k_BT$ (on the order of a few microns at room temperature)--this exchange is bounded by the Stefan–Boltzmann law. However, as first demonstrated by Rytov and later formalized by Polder and van Hove~\cite{Rytov53,PoldervH}, this limit breaks down in the near-field regime, where the separation falls below  $\lambda_\mathrm{th}$. These pioneer works were followed by a plethora of theoretical and experimental developments (see Refs.~\cite{Volokitin07,Song15,Cuevas18,Biehs21} and references therein). In the near field, the radiative heat flux can be dramatically enhanced due to the tunneling of evanescent photons. This enhancement becomes especially pronounced when the interacting bodies support surface resonances such as phonon-polaritons in polar dielectrics or surface plasmons in metals~\cite{Joulain05}, and can be further amplified by broadband evanescent modes in hyperbolic materials~\cite{Biehs12}. These phenomena have opened promising avenues for nanoscale thermal management~\cite{Latella21a}, solid-state cooling~\cite{Chen15,Zhu19}, infrared sensing and spectroscopy~\cite{De Wilde06,Jones12}, energy conversion~\cite{DiMatteo01,Narayanaswamy03,Laroche06,Park08,Latella21b}, and the emerging field of thermotronics~\cite{BenAbdallah13a,BenAbdallah16PRB,BenAbdallah15,Reddy24}.

An additional application domain is heat-assisted magnetic recording (HAMR)~\cite{Srituravanich04,Challener09,Stipe10}, where localized heating is used to transiently lower the coercivity of a magnetic substrate, enabling data writing with an external magnetic field. Achieving high information densities requires the confinement of heat to nanoscale regions, ideally approaching the superparamagnetic limit of tens of nanometers. In this context, we study in the present work the spatially resolved near-field radiative heat transfer between a nanoparticle and a substrate in the fluctuational-electrodynamics framework. Inspired by HAMR-like configurations,  we address in this work both key aspects--enhanced heat transfer and spatial localization--by investigating near-field radiative heat exchange between a nanoparticle and a planar substrate. The theoretical framework employed falls within the well-established domain of dipolar near-field heat transfer, which has garnered significant interest over the past decade~\cite{BenAbdallah11,Messina13,BenAbdallah13b,Tervo19,Luo20,Fang23,Biehs16,Dong18,Messina18,Deshmukh18,Zhang19a,Zhang19b,Ott20,Zhang20,Ott21,Fang22,Asheichyk22,Saaskilahti14,Asheichyk18,Zhang21,Chen22,Zhang23,Asheichyk,BenAbdallah19,Rihouey25}. Motivated by the HAMR concept, we consider a configuration where a nanoparticle--serving as a localized heat source--is positioned near a multilayer substrate composed of different materials. It is shown that the introduction of a thin polar film atop a non-dispersive substrate can lead to a substantial enhancement and spatial focusing of the radiative heat flux.

The paper is structured as follows. In Sec.~\ref{sec:system} we present the physical system and the employed technical tools. The results are presented in Sec.~\ref{sec:results}. First, Sec.~\ref{sec:results1} demonstrates the possibility to enhance and focus the particle--substrate heat flux by modulating the film thickness. Section~\ref{sec:results2} provides a spectral interpretation of the effect by analyzing the dispersion relations of electromagnetic resonant modes, while Sec.~\ref{sec:results3} explores the dependence on the particle--substrate distance, highlighting the near-field nature of the effect. The last Section presents some conclusive remarks and perspectives.

\section{Physical system}\label{sec:system}

\begin{figure}
	\centering
	\includegraphics[width=0.35\textwidth]{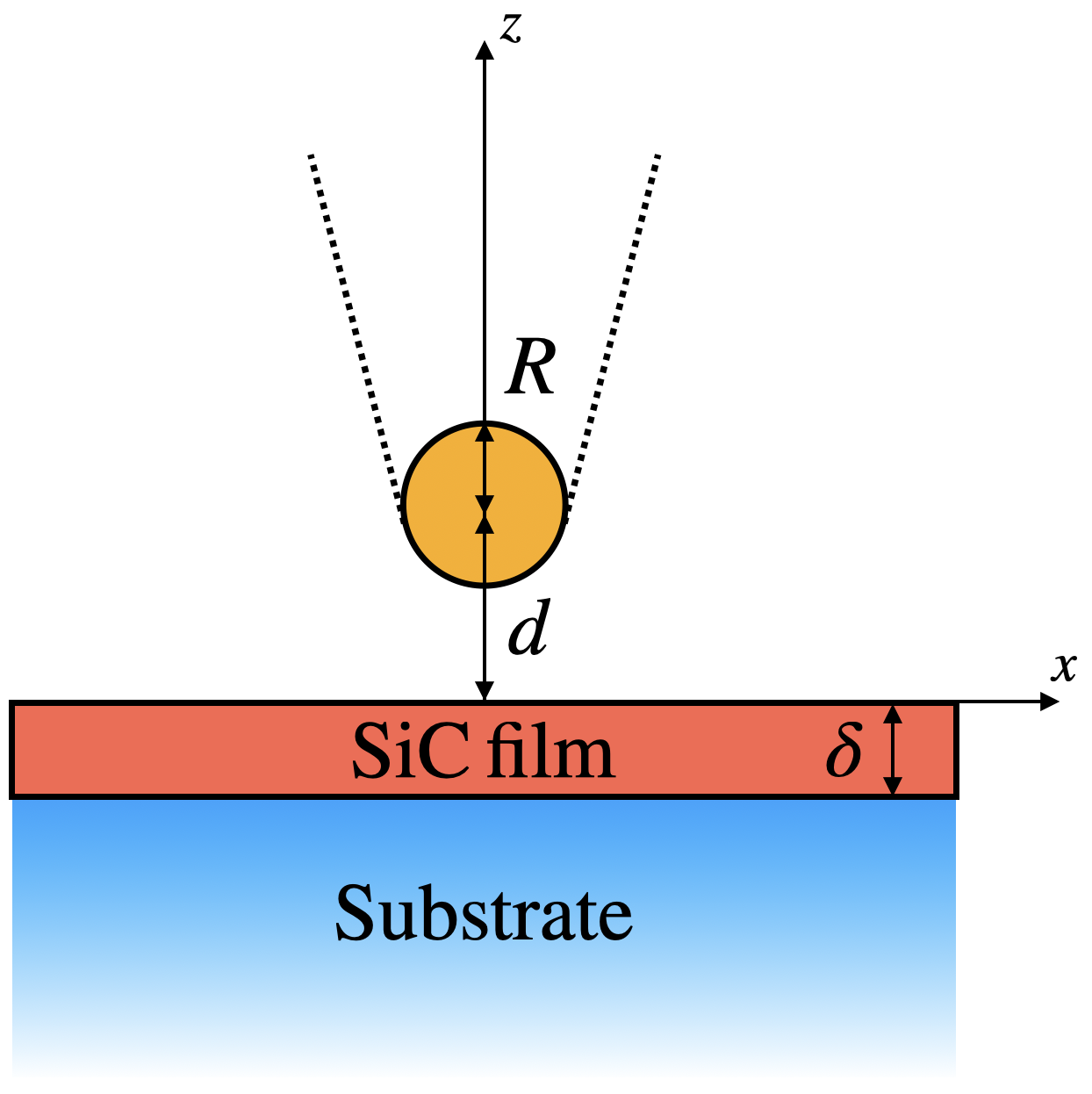}
	\caption{Geometry of the system. A small sphere of radius $R$, mimicking the presence of a tip, is placed at distance $d$ from a substrate covered by a SiC film of thickness $\delta$. The limiting case $\delta=0$ corresponds to the absence of film. Both the substrate and the film are translationally invariant along the $x$ and $y$ axes.}\label{Fig:Geometry}
\end{figure}

The system we consider, sketched in Fig.~\ref{Fig:Geometry}, is made of a spherical nanoparticle of radius $R$ simulating a tip, made of silicon carbide (SiC). The nanosphere is placed at a distance $d$ (along the $z$ axis) from a substrate covered by a film of thickness $\delta$ made of SiC, both assumed to be infinite along the $x$ and $y$ directions. In the following, the optical properties of SiC will be described by a Drude-Lorentz model~\cite{Palik}
\begin{equation}
	\varepsilon(\omega)=\varepsilon_\infty\frac{\omega^2_L-\omega^2-i\Gamma\omega}{\omega^2_T-\omega^2-i\Gamma\omega},\end{equation}
where $\varepsilon_\infty=6.7$, $\omega_L=1.827\times 10^{14}\,$rad/s, $\omega_T=1.495\times 10^{14}\,$rad/s, and $\Gamma=9\times 10^{11}\,$rad/s. Concerning the underlying substrate, we are going to explore a variety of non-dispersive materials having permittivities $\varepsilon_\text{sub}=2,4,8,16$ and a very small imaginary part included for numerical reasons. The reason for this choice of permittivites is twofold. On one hand, sweeping over a range of possible permittivites allows us to investigate on the robustness of the effect associated with the thin film. On the other hand, the optical properties of the magnetic materials typically employed in HAMR are currently scarcely available in the far-infrared region of the spectrum, the most relevant one for radiative heat transfer close to ambient temperature. One result in this sense is the measurement of the refractive index of FePt thin films~\cite{FePt}, resulting in a value of $n\simeq 4$ at $\lambda=1.7\,\mu$m, motivating our choice of maximum $\varepsilon_\text{sub}$ of 16.

In this work we are going to focus on the spatial distribution of the energy transferred between the nanosphere and the underlying substrate. This is encoded in the $z$ component $S_z$ of the Poynting vector $\mathbf{S}=\mathbf{E}\times\mathbf{H}$, to be evaluated at the interface $z=0$ between vacuum and substrate. This quantity can be calculated within a $N$-body fluctuational-electrodynamics approach complemented with the dipolar approximation, discussed e.g. in Ref.~\cite{Rihouey25}. In this framework, each nanoparticle is described as an electric dipole, sum of a fluctuating part due to its thermal fluctuations and an induced one stemming from its response to the external field, coming from an external thermal bath and the other dipoles. The total field, and thus the Poynting vector, results from the use of the fluctuation-dissipation theorem and the knowledge of the system Green's function.

In our scenario, we are going to focus on a single nanoparticle at temperature $T=400\,$K, and assume that both the thermal bath and the substrate are at zero temperature. This allows us to focus purely on the contribution to the energy arriving on the substrate coming from the nanoparticle thermal emission. In this case the $\alpha$ component ($\alpha=x,y,z$) of the Poynting vector at position $\mathbf{R}$ will be written as a spectral decomposition
\begin{equation}
S_\alpha(\mathbf{R}) = \int_0^{\infty}d\omega\,S_\alpha(\mathbf{R},\omega),
\end{equation}
where the spectral component reads~\cite{Rihouey25}
\begin{equation}\begin{split}
		S_\alpha(\mathbf{R},\omega) &= -\frac{4\hbar\omega^4}{c^3}\chi(\omega)n(\omega,T_d)\epsilon_{\alpha\beta\gamma}\\
		&\,\times\mathrm{Im}\Bigl[\mathds{G}^{\text{(EE)}}_{\beta\beta'}(\mathbf{R},\mathbf{R}_d)\mathds{G}^{\text{(HE)}*}_{\gamma\gamma'}(\mathbf{R},\mathbf{R}_d)\Bigr].\end{split}\label{eq:Salpha}\end{equation}
In this expression $\epsilon_{\alpha\beta\gamma}$ is the Levi-Civita tensor, 
\begin{equation}
	n(\omega,T) = \biggl[\exp\biggl(\frac{\hbar\omega}{k_B T}\biggr) - 1\biggr]^{-1}\end{equation}
the Bose-Einstein distribution at temperature $T$, while $\chi(\omega)$ is the nanoparticle susceptibility, which can be expressed as~\cite{Manjavacas12,Messina13}
\begin{equation}
	\chi(\omega) = \mathrm{Im}[\alpha(\omega)]-\frac{\omega^3}{6\pi c^3}|\alpha(\omega)|^2,
\end{equation}
in terms of the polarizability $\alpha(\omega)$, expressed here by means of the Clausius-Mossotti relation
\begin{equation}
 \alpha(\omega) = 4\pi R^3\frac{\varepsilon(\omega)-1}{\varepsilon(\omega)+2}.
\label{eq:alpha}\end{equation}
Moreover, Eq.~\eqref{eq:Salpha} contains the electric-electric and magnetic-electric Green's function of the system calculated at position $\mathbf{R}$ and for a given particle position $\mathbf{R}_d$. In the presence of a scatterer (the film-covered substrate), these functions can be expressed as the sum of a vacuum and a scattering contributions. The former read
\begin{equation}\begin{split}
		\mathds{G}^{(0)}_\text{EE}(\omega,\mathbf{R},\mathbf{R}')&=\frac{e^{ik_0d}}{4\pi d}\Bigl[\Bigl(1+\frac{ik_0d-1}{k_0^2d^2}\Bigr)\mathds{I}\\
		&\,+\frac{3-3ik_0d-k_0^2d^2}{k_0^2d^2}\hat{\mathbf{d}}\otimes\hat{\mathbf{d}}\Bigr],\\
		\mathds{G}^{(0)}_\text{HE}(\omega,\mathbf{R},\mathbf{R}')&=\frac{e^{ik_0d}}{4\pi d}\frac{ik_0d-1}{k_0d^2}\begin{pmatrix}0 & -d_z & d_y\\d_z & 0 & -d_x\\-d_y & d_x & 0\end{pmatrix},
\end{split}\end{equation}
were $k_0=\omega/c$ and we have introduced the distance $\mathbf{d} = \mathbf{R} - \mathbf{R}'$ having norm $d=|\mathbf{d}|$ and such that $\hat{\mathbf{d}}=\mathbf{d}/d$ and
\begin{equation}
	\mathbf{d} = d(\sin\theta_d\cos\varphi_d,\sin\theta_d\sin\varphi_d,\cos\theta_d).
\end{equation}
The scattering contribution to the Green's function reads~\cite{Novotny} for $z,z'>0$ (both arguments $\mathbf{R}$ and $\mathbf{R}'$ above the substrate)
\begin{equation}\begin{split}
		\mathds{G}_\text{EE}&(\omega,\mathbf{R},\mathbf{R}')=\int\frac{dk}{2\pi}\frac{ike^{i  k_z(z+z')}}{2k_z}\Bigl[r_\text{TE}\begin{pmatrix}A & C & 0\\C & B & 0\\0 & 0 & 0\end{pmatrix}\\
		&+r_\text{TM}\frac{c^2}{\omega^2}\begin{pmatrix}-k_z^2B & k_z^2C & -  kk_zE\\k_z^2C & -k_z^2A & -  kk_z D\\  k k_zE & 
			  kk_z D& k^2F\end{pmatrix}\Bigr],\\
		\mathds{G}_\text{HE}&(\omega,\mathbf{R},\mathbf{R}')=\int\frac{dk}{2\pi}\frac{cke^{i  k_z(z+z')}}{2\omega k_z}\\
		&\times\Bigl[r_\text{TE}\begin{pmatrix}  k_z C &   k_z B & 0\\-  k_z A & -  k_z C & 0\\k D & - k E & 0\end{pmatrix}\\
		&-r_\text{TM}\begin{pmatrix}-  k_z C &   k_z A & k D\\-  k_z B &   k_z C & -k E\\0 & 0 & 0\end{pmatrix}\Bigr],
\end{split}\label{eq:Green}\end{equation}
where
\begin{equation}
	\begin{pmatrix}A\\B\\C\\D\\E\\F\end{pmatrix}=\begin{pmatrix}\frac{1}{2}[J_0(kd) + J_2(kd)\cos(2\varphi_d)]\\ \frac{1}{2}[J_0(kd) - J_2(kd)\cos(2\varphi_d)]\\\frac{1}{2}J_2(kd)\sin(2\varphi_d)\\i J_1(kd)\sin\varphi_d\\ i J_1(kd)\cos\varphi_d\\J_0(kd)\end{pmatrix}.\end{equation}

\section{Numerical results}\label{sec:results}

\subsection{Poynting vector amplification and focusing}\label{sec:results1}

For our first numerical simulations, we start from a nanoparticle having radius $R=100\,$nm. It has been shown~\cite{BenAbdallah08} that the validity of the dipolar approximation is limited to configurations in which the distance between the particle and any adjacent body is at least three times the radius. For this reason, and in order to maximize near-field effects, we pick here the minimum allowed distance $d=3R=300\,$nm. Our objective is to analyze the impact of the film and more specifically of its thickness on the amplification and focusing of energy transfer with respect to the reference configuration corresponding to the absence of film (substrate only). For each value of $\delta$, the former will be quantitavely done by studying the ratio $S_z(0,0,0)/S_\mathrm{ref}$ between the $z$ component of the Poynting vector right below the particle (on the vacuum side of the interface) and the reference value corresponding to the scenario in the absence of film, i.e. $\delta=0$. As far as the focusing is concerned, it will be described by the half width at half maximum (HWHM) of the profile of $S_z(x,y,0)$, i.e. the $x$ coordinate (in virtue of the rotational symmetry of the system) at which $S_z(x,0,0)=S_z(0,0,0)/2$.

\begin{figure}[t!]
	\centering
	\includegraphics[width=0.45\textwidth]{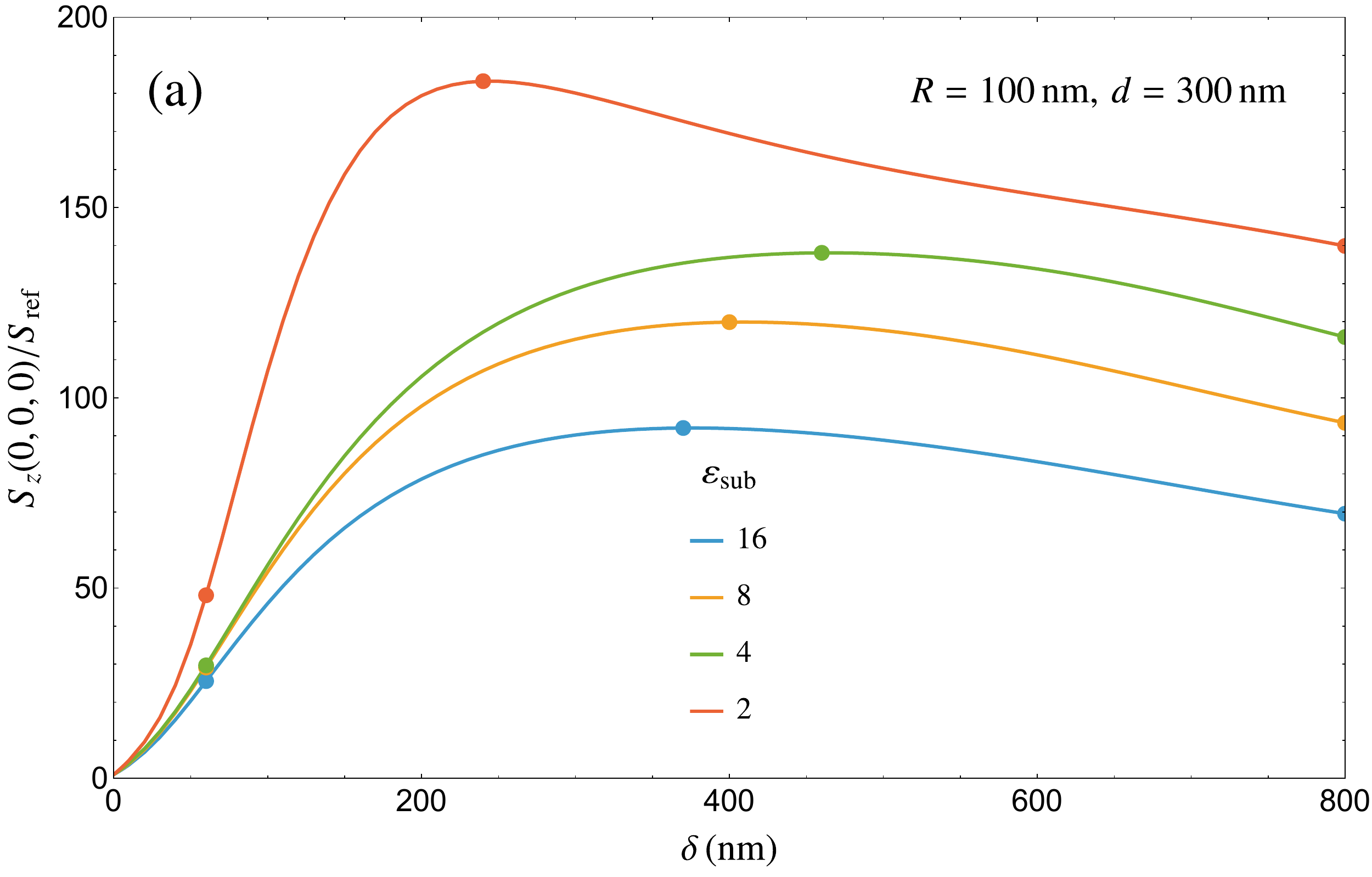}
	\includegraphics[width=0.45\textwidth]{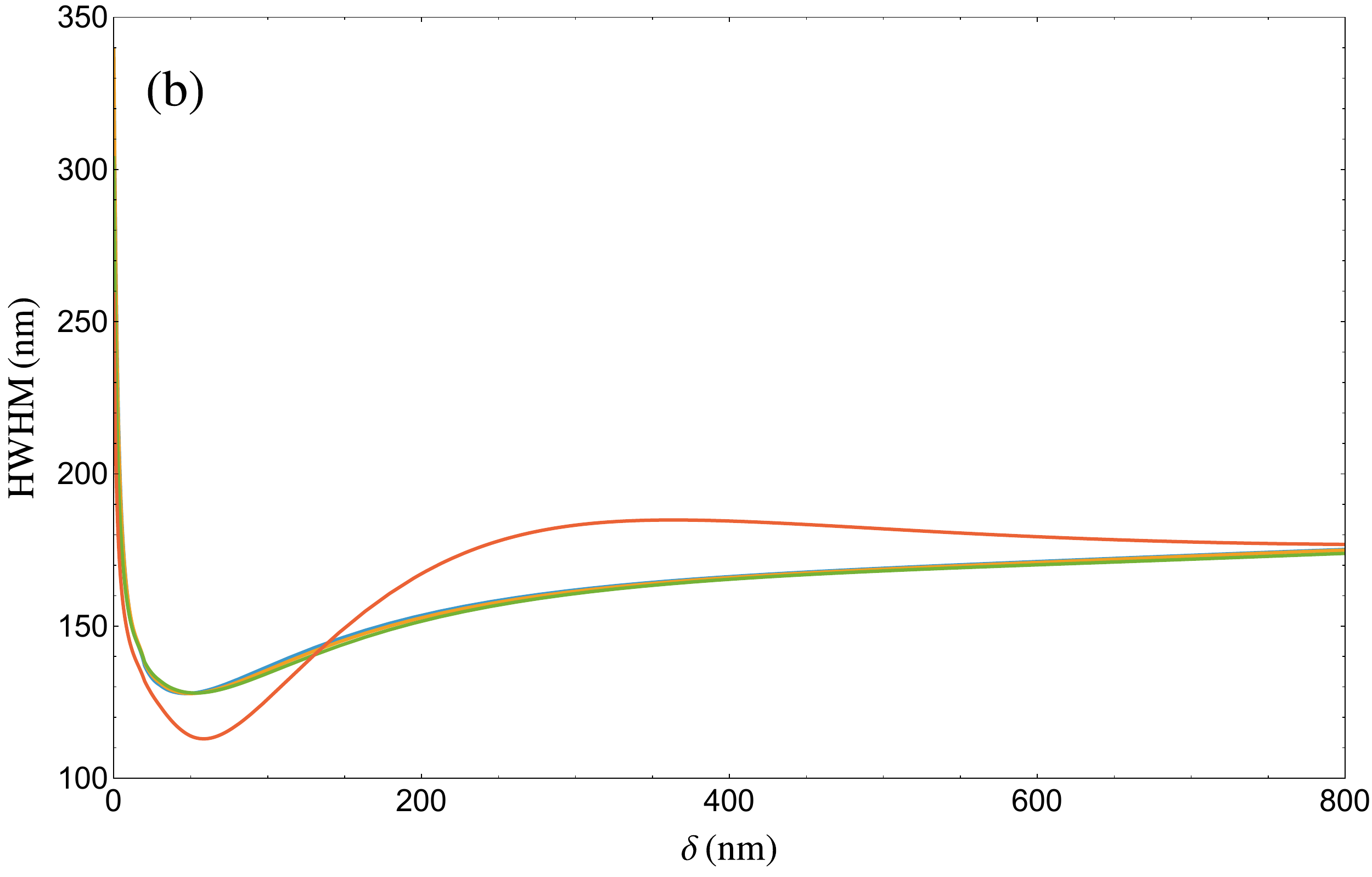}
	\caption{(a) Ratio of the Poynting vector $S_z(0,0,0)$ below the nanoparticle and the one $S_\mathrm{ref}$ in the absence of film ($\delta=0$) as a function of the film thickness $\delta$. The 4 curves correspond to different values of the substrate permittivity (see legend). (b) Corresponding values of the HWHM.}\label{Fig:Main}
\end{figure}

Figure~\ref{Fig:Main} shows these two quantites as a function of the film thickness $\delta$ for four different values of the substrate permittivity $\varepsilon_\mathrm{sub}=2,4,8,16$. In Fig.~\ref{Fig:Main}(a), we see that for any chosen substrate permittivity, the curve starts at 1 as expected and shows a non-monotonic behavior with respect to $\delta$, with a pronounced peak varying between 92 (for $\varepsilon_\mathrm{sub}=16$) and 183 (for $\varepsilon_\mathrm{sub}=2$). This first result already shows how a thin film having a thickness of hundreds of nanometers can induce a two-order-of-magnitude increase of the energy transferred between a nanoparticle and a substrate. Figure~\ref{Fig:Main}(b) shows that the presence of the thin film also has a strong impact on the width of the Poynting vector profile. These curves are also non-monotonic, with a FWHM going from the values $340, 340, 304, 260\,$nm for $\varepsilon_\mathrm{sub}=2,4,8$ and 16, respectively, in the absence of film to 113, 128, 128, 128\,nm in the best scenario in the presence of a film.

\begin{figure}
	\centering
	\includegraphics[width=0.45\textwidth]{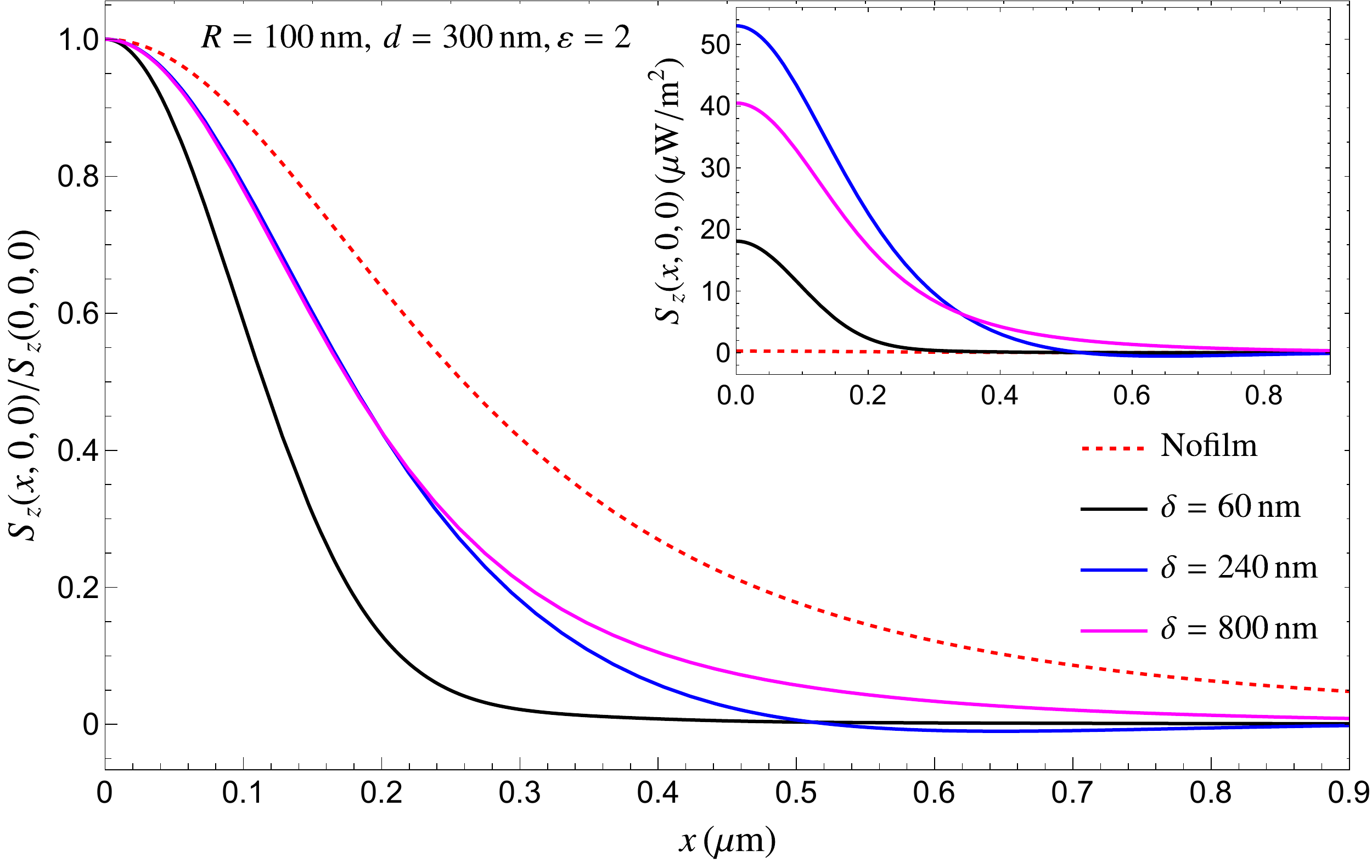}
	\caption{Profile of the Poynting vector $S_z(x,0,0)$ in normalized (main part) and absolute (inset) units for $\varepsilon_\mathrm{sub}=2$ and $\delta=0,60,240,800\,$nm (see legend).}\label{Fig:Profile}
\end{figure}

This view can be completed by looking at $S_z(x,0,0)$ as a function of $x$ for $\varepsilon_\mathrm{sub}=2$ and $\delta=60,240,800\,$nm. These three representative values of the thickness are chosen to be the one at which the HWHM is at its minimum, the one maximizing $S_z(0,0,0)$ and the largest one in Fig.~\ref{Fig:Main}. They are highlighted in Fig.~\ref{Fig:Main}(a) for all $\varepsilon_\mathrm{sub}$. Note that the $\delta$ maximizing $S_z(0,0,0)$ depends on $\varepsilon_\mathrm{sub}$. The profiles for $\varepsilon_\mathrm{sub}=2$ and $\delta=60,240,800\,$nm are shown in Fig.~\ref{Fig:Profile} along with the one in the absence of film ($\delta=0$) in normalized (main part) and absolute (inset) units. The inset of the figure confirms the strong amplification for the intermediate value of $\delta$, while the main part of the figure highlights clearly the optimized HWHM.

\subsection{Spectral analysis and dispersion relations}\label{sec:results2}

We now turn to the interpretation of the effect highlighted in the previous section. To this aim we focus on the spectral analysis of the Poynting vector for one specific value of the substrate permittivity, namely $\varepsilon_\mathrm{sub}=2$, and for the value of thickess $\delta$ mentioned before, highlighted by the points in Fig.~\ref{Fig:Main}(a) and shown in Fig.~\ref{Fig:Profile}. Before looking at the results, it is useful to review some important and expected spectral feature in the context of near-field radiative heat transfer. As already discussed in the introduction, it is well known that near-field heat transfer amplification is mainly due to the existence of resonant modes of the electromagnetic field propagating along the interface between a body and vacuum and exponentially decaying when getting far from it. Such resonant modes in the case of a nanosphere correspond to resonances of its polarizability. From its definition in Eq.~\eqref{eq:alpha}, we see that this corresponds to $\varepsilon(\omega)=-2$, which for our model of SiC gives $\omega\simeq1.756\times10^{14}$\,rad/s. Since we have chosen a non-dispersive model for the substrate permittivity, no resonant modes are expected in the absence of film. On the contrary, the scenario is expected to change in the presence of a film of SiC, possibly supporting resonant modes. More specifically, while a SiC semi-infinite substrate supports a mode resonating at the frequency such that $\varepsilon(\omega)=-1$, the discussion becomes more involved for a finite $\delta$, as shown in the following.

\begin{figure}
	\centering
	\includegraphics[width=0.45\textwidth]{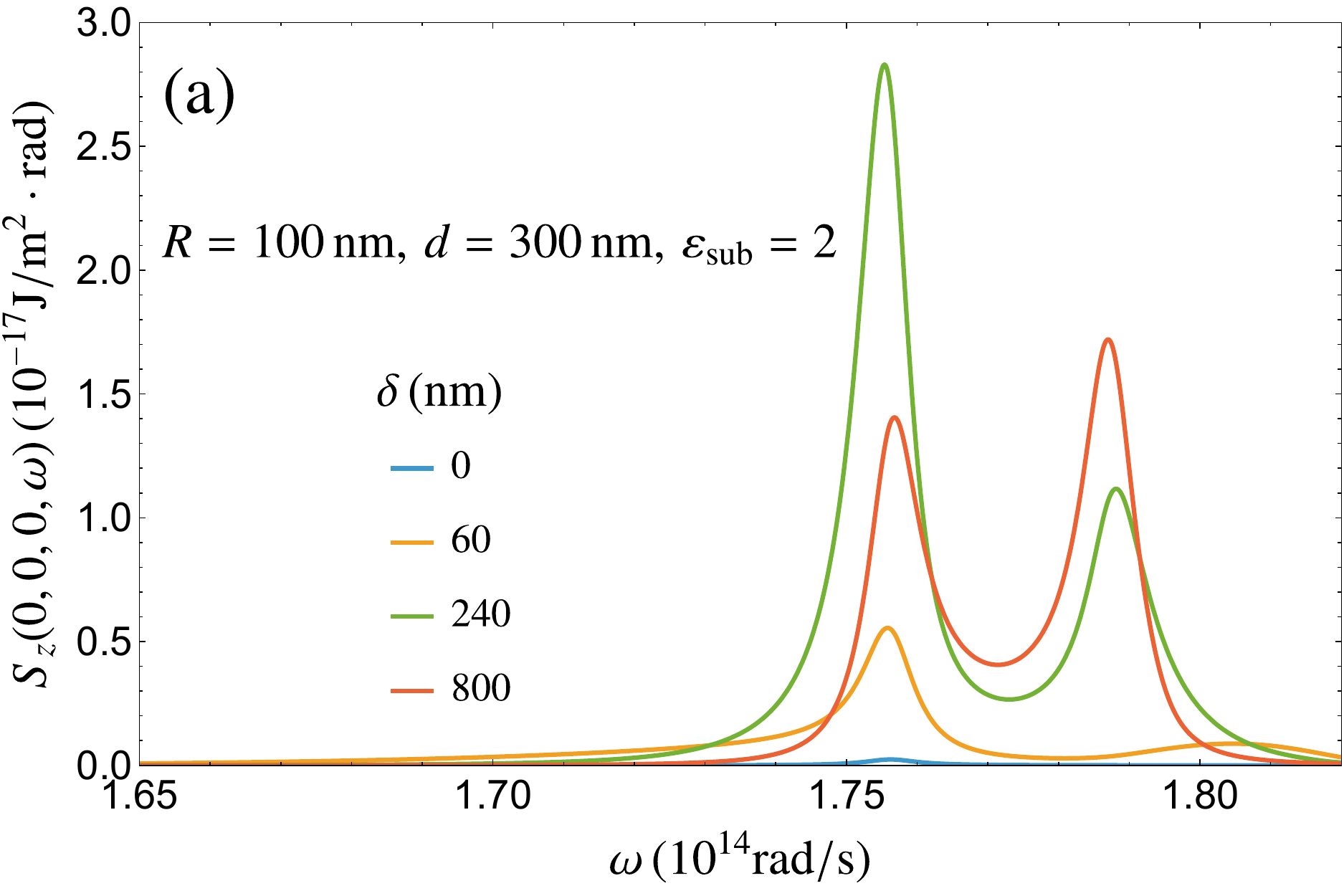}
	\includegraphics[width=0.45\textwidth]{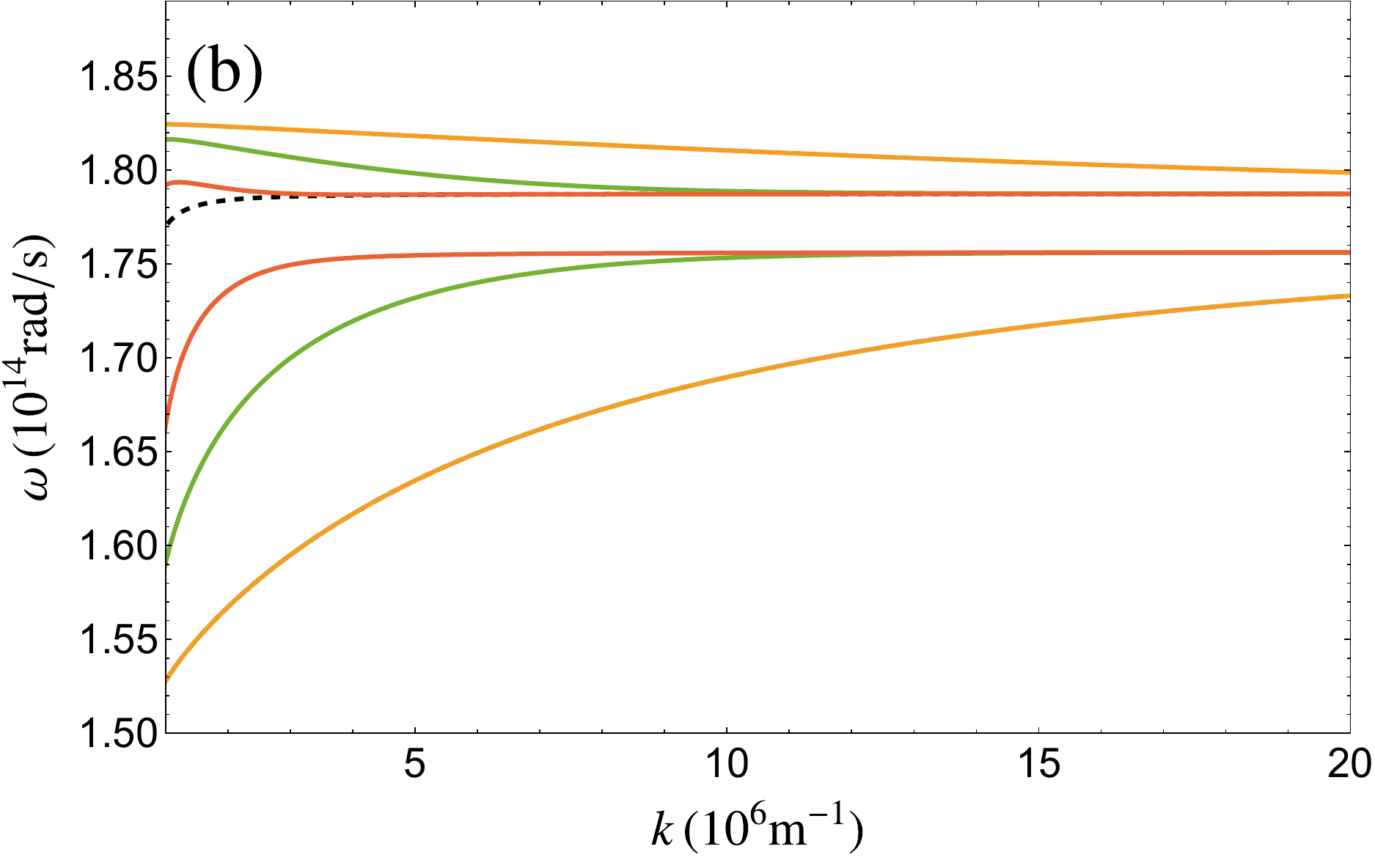}
	\caption{(a) Spectrum of the Poynting vector at the origin (below the particle) for different values of film thickness and in the absence of film (see legend). (b) Dispersion relation of the surface modes for the vacuum--film--substrate system, along with the one for a SiC-vacuum interface (black dashed line).}\label{Fig:Spectrum}
\end{figure}

Figure~\ref{Fig:Spectrum}(a) shows the spectral Poynting vector in our reference configuration ($R=100\,$nm, $d=300\,$nm, $\varepsilon_\mathrm{sub} =2$) in the absence of film and for the 3 chosen thicknesses. We remark, as expected, the presence of a clear peak at the nanoparticle resonance $\omega\simeq1.756\times10^{14}$\,rad/s. This peak is the only one visible in the absence of film, and its height is sensitive to the chosen thickness. Moreover, for the three non-vanishing values of $\delta$, a second peak is visible, and its position clearly depends on $\delta$. Such behavior can be interpreted by analyzing the dispersion relation of the surface modes supported by the finite-thickness film on top of the non-dispersive substrate. This can be done by solving Maxwell's equations in the three regions and imposing the continuity conditions at each interface. As a result, it can be shown analytically that the dispersion relation $\omega(k)$ of the resonant mode stems from solution of the equation~\cite{Economou69}
\begin{equation}\begin{split}
	&\hspace{-2cm}(\varepsilon_\mathrm{film}(\omega)k_z + k_{z,\mathrm{film}})(\varepsilon_\mathrm{sub}k_{z,\mathrm{film}}+\varepsilon_\mathrm{film}(\omega)k_{z,\mathrm{sub}})\\
	\,-e^{2ik_{z,\mathrm{film}}\delta}&(-\varepsilon_\mathrm{film}(\omega)k_z + k_{z,\mathrm{film}})\\
	\times&(\varepsilon_\mathrm{sub}k_{z,\mathrm{film}}-\varepsilon_\mathrm{film}(\omega)k_{z,\mathrm{sub}})=0,
\end{split}\label{eq:dispersion}\end{equation}
where the $z$ components of the wavevector in the three media read
\begin{equation}\begin{split}
	&k_z=\sqrt{\frac{\omega^2}{c^2}-k^2},k_{z,\mathrm{film}}=\sqrt{\varepsilon_\mathrm{film}(\omega)\frac{\omega^2}{c^2}-k^2},\\
	&\hspace{2cm}k_{z,\mathrm{sub}}=\sqrt{\varepsilon_\mathrm{sub}\frac{\omega^2}{c^2}-k^2}.
\end{split}\end{equation}
Two important features of Eq.~\eqref{eq:dispersion} are worth underlying for our purposes. The former is the explicit dependence on $\delta$ through the exponential factor. The latter is that in the limit of large wavevectors this exponential factor vanishes, we have $k_z,k_{z,\mathrm{film}},k_{z,\mathrm{sub}}\to ik$ and the equation simplifies to
\begin{equation}(\varepsilon_\mathrm{film}(\omega) + 1)(\varepsilon_\mathrm{sub}+\varepsilon_\mathrm{film})=0,\end{equation}
leading to the condition $\varepsilon_\mathrm{film}(\omega) = - 1$, known for a semi-infinite substrate, and the new one $\varepsilon_\mathrm{film}=-\varepsilon_\mathrm{sub}$, stemming from the second interface and dependent on $\varepsilon_\mathrm{sub}$.

We have solved Eq.~\eqref{eq:dispersion} for real $k$ and complex $\omega$ and plotted the real part of $\omega$ as a function of $k$ in Fig.~\ref{Fig:Spectrum}(b). As a reference, the black dashed curve shows the case of a semi-infinite SiC substrate. In this case, the horizontal asymptote corresponds to the solution of $\varepsilon_\mathrm{film}(\omega) = - 1$, as expected. On the contrary, in the case of a film between vacuum and a substrate  the dispersion relation is split into two branches (commonly referred to as symmetric and anti-symmetric because of the associated electric-field distribution), one below and one above the one for a SiC substrate. While the one above tends to the solution of $\varepsilon_\mathrm{film}(\omega) = - 1$, the one below goes to the frequency such that $\varepsilon_\mathrm{film}=-\varepsilon_\mathrm{sub}$. Note also that the slope at which these branches tend to their asymptotic value strongly depends on $\delta$. Moreover, we also have to keep in mind that as a general rule in near-field radiative heat transfer, the distance between bodies exchanging heat dictates the range of wavevectors participating to the energy exchange. When the geometry under scrutiny allows for an analytical expression of the flux, this feature is typically encoded in an exponential factor $\exp(-\gamma k d)$, which in our case appears inside the scattering contribution to the Green's function given in Eq.~\eqref{eq:Green}. This explains, along with the analysis of the three upper branches in Fig.~\ref{Fig:Spectrum}(b), the behavior of the upper-frequency peaks in Fig.~\ref{Fig:Spectrum}(a). For $\delta=60\,$nm the dispersion relation does not clearly reach its asymptote within the considered range of $k$, explaining the lower peak in the spectrum, appearing at higher frequencies. On the contrary, switching to 240\,nm and 800\,nm, the asymptote is reached earlier with respect to $k$, thus producing a higher peak in the spectrum. The analysis of the lower peaks steams insted from the coupling between the nanoparticle resonance and the dispersion relations. For $\delta=60\,$nm the asymptote is not reached in the considered range, producing only a small increase in the peak. On the contrary, the opposite takes place for higher thicknesses, explaining the enhanced peak, and also the enlargement seen for $\delta=240\,$nm. Note that this analysis would change for a different value of $\varepsilon_\mathrm{sub}$ since this would have an impact on the asymptote of the lower branches.

While the discussion has been done for specific values of $d$, $\delta$ and $\varepsilon_\mathrm{sub}$, the general features and the main message clearly emerged. The possibility of enhancing the heat transferred between the nanoparticle and the substrate thanks to the presence of a thin film mainly stems from the appearance of new resonant modes at the vacuum--film--substrate interfaces. These strongly depend on the substrate permittivity (dictating one of the asymptotes of the dispersion relation) and on the film thickness, having a strong impact on the shape of both branches. These shapes, in conjunction with the particle--substrate distance, have an influence on the field modes participating to the effect and thus on the flux amplification. Concerning the reduction of HWHM, its origin has been already discussed in Ref.~\cite{Rihouey25}, where this effect was studied in the presence of magneto-optical nanoparticles. As a matter of fact, it was shown that near-field interactions can induce an amplification of Poynting vector which happens all over the plane, and not only below the particle. Nevertheless, the near-field nature of this effect is such that this amplification is more pronounced in the closest regions, i.e. at the origin, naturally leading to a more focused energy flux.

\subsection{Impact of particle-substrate distance}\label{sec:results3}

\begin{figure}
	\centering
	\includegraphics[width=0.45\textwidth]{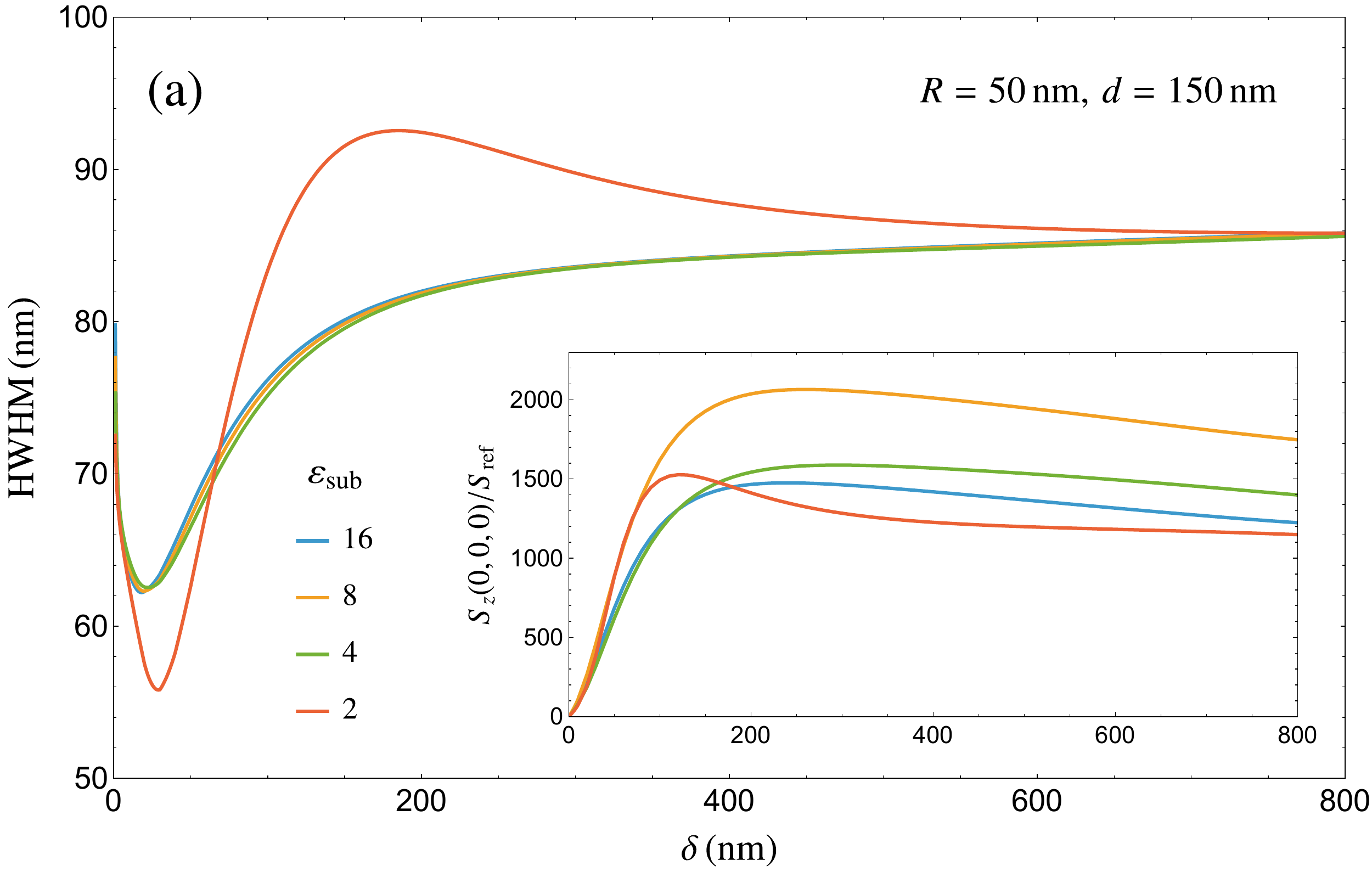}
	\includegraphics[width=0.45\textwidth]{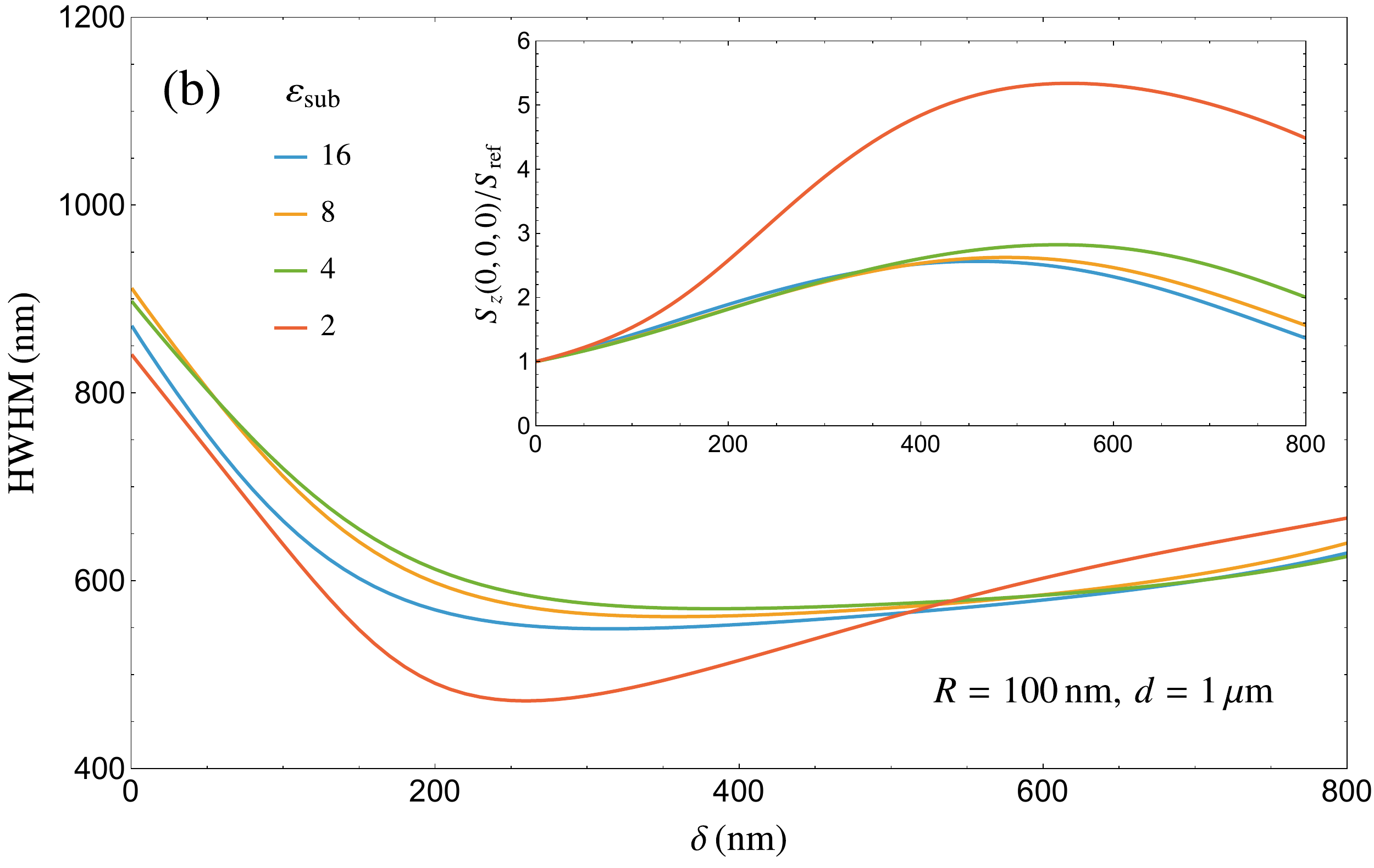}
	\caption{HWHM (main part) and Poynting vector amplification (inset) for (a) $R=50\,$nm and $d=150\,$nm, (b) $R=100\,$nm and $d=1\,\mu$m.}\label{Figd:d1501000}
\end{figure}

We finally want to show explicitly that the observed effect stems from near-field interactions. To this aim we consider two different scenarios, one in which we consider a smaller radius $R=50\,$nm, allowing us to explore a smaller distance $d=150\,$nm, and one where for the initially chosen radius $R=100$\,nm we move to larger distance $d=1\,\mu$m. Figure~\ref{Figd:d1501000} shows the results, leading to the observation of the same qualitative behavior of these two further configurations compared to the initial one. For both we observe a heat flux amplification and a reduction of the HWHM. Nevertheless, the factors of flux amplification vary dramatically, with an enhancement reaching a factor 2000 for $d=150\,$nm, and going down to 6 for $d=1\,\mu$m. We also note that the film thickness at which the extreme values flux amplification and HWHM reduction are realized strongly depend on $d$. This shows unambiguously that we are indeed in the presence of a near-field effect and thus that the particle--substrate distance plays a major role.

\section{Conclusions}

Using a fluctuational-electrodynamics framework within the dipolar approximation, we have demonstrated that the spatially resolved near-field radiative heat transfer between a polar nanoparticle and a non-dispersive substrate can be significantly controlled by introducing a thin polar film atop the substrate. By analyzing the magnitude and spatial distribution of the Poynting vector, we have shown that specific film thicknesses can lead to a pronounced enhancement and focusing of the local energy flux.

These effects are attributed to the interplay between the dispersion relations of resonant electromagnetic modes supported by the vacuum–film–substrate configuration and the nanoparticle–substrate separation. We have highlighted that the dispersion characteristics are strongly influenced by the film thickness and substrate permittivity, while the particle–substrate distance determines the wavevectors contributing most significantly to the heat exchange. As a result, the optimization of heat transfer enhancement and spatial confinement depends critically on the careful selection of geometric and material parameters, constrained by both physical feasibility and fabrication limitations.

Furthermore, our analysis has revealed that the conditions which maximize heat flux and minimize the half-width at half-maximum of the spatial profile generally occur at different film thicknesses. This suggests that, for practical applications, an appropriate scalar objective function must be defined—balancing both enhancement and localization—so that the presented methods can be employed to optimize system performance accordingly.

Overall, our findings provide insights into the control of spatially resolved thermal radiation and point toward new strategies for nanoscale thermal management. These results are particularly relevant for applications such as heat-assisted magnetic recording and may be extended to more complex scenarios, including multi-tip geometries and self-consistent treatment of heat diffusion within multilayer substrates.

\end{document}